\documentclass[12pt,preprint]{aastex}

\def\spose#1{\hbox to 0pt{#1\h\> \> SS}}
\def\simlt{\mathrel{\spose{\lower 3pt\hbox{$\mathchar"218$}}
     \raise 2.0pt\hbox{$\mathchar"13C$}}}
\def\simgt{\mathrel{\spose{\lower 3pt\hbox{$\mathchar"218$}}
     \raise 2.0pt\hbox{$\mathchar"13E$}}}

\def\mic{{$\mu$m}}

\def\h2o{H$_2$O}

\def\aple{$\mathrel{\hbox{\rlap{\hbox{\lower4pt\hbox{$\sim$}}}\hbox{$<$}}}$}

\def\apge{$\mathrel{\hbox{\rlap{\hbox{\lower4pt\hbox{$\sim$}}}\hbox{$>$}}}$}

\slugcomment{}
\lefthead{Blum {\it et al.}}
\righthead{SAGE Evolved Stars}

\begin{document}
\title{Spitzer SAGE survey of the Large Magellanic Cloud II: Evolved
Stars and Infrared Color Magnitude Diagrams}

\author{R. D. Blum\footnote{Cerro Tololo Interamerican Observatory,
Casilla 603, La Serena, Chile} \ , J. R. Mould\footnote{NOAO, PO Box
26732, Tucson AZ 85726-6732} \ , K. A. Olsen$^1$,
J. A. Frogel\footnote{AURA, Inc., 1200 New York Ave. NW, Suite 350,
Washington D.C. 20005} \ , M. Werner\footnote{Jet Propulsion Lab, 4800
oak Grove Dr., MS 264--767, Pasadena, CA 91109} \ ,
M. Meixner\footnote{Space Telescope Science Institute, 3700 San Martin
Way, Baltimore, MD 21218} \ , F. Markwick--Kemper\footnote{Department
of Astronomy, University of Virginia, PO Box 3818, Charlottesville, VA
22903} \ , R. Indebetouw$^6$, B. Whitney\footnote{Space Science
Institute, 308 Morningside Ave., Madison, WI 53716} \ ,
M. Meade\footnote{Department of Astronomy, 475 North Charter St.,
University of Wisconsin, Madison, WI 53706} \ , B. Babler$^{8}$,
E. B. Churchwell$^8$, K. Gordon\footnote{Steward Observatory,
University of Arizona, 933 North Cherry Ave., Tucson, AZ 85719} \ ,
C. Engelbracht$^9$, B--Q For$^9$, K. Misselt$^9$, U. Vijh$^5$,
C. Leitherer$^5$, K. Volk\footnote{Gemini Observatory, 670 North
A'ohuku Place, Hilo, HI 96720} \ , S. Points$^1$,
W. Reach\footnote{Spitzer Science Center, California Institute of
Technology, 220-6, Pasadena, CA, 91125} \ , J. L. Hora\footnote{Center
for Astrophysics, 60 Garden St., MS 67 , Harvard University,
Cambridge, MA 02138} \ , J--P. Bernard\footnote{Centre d' \'{E}tude
Spatiale des Rayonnements, CNRS, 9 av. du Colonel Roche, BP 4346,
31028 Toulouse, France} \ , F.  Boulanger\footnote{Institut
d'Astrophysique Spatiale, Universit\'{e} Paris-XI, 91405 Orsay Cedex,
France} \ , S. Bracker$^8$, M. Cohen\footnote{Radio Astronomy
Laboratory, 601 Campbell Hall, University of California at Berkeley,
Berkeley, CA 94720} \ , Y. Fukui\footnote{Department of Astrophysics,
Nagoya University, Chikusa-ku, Nagoya 464-8602, Japan} \ ,
J. Gallagher$^{8}$, V. Gorjian$^{11}$, J. Harris$^9$, D. Kelly$^9$,
A. Kawamura$^{16}$, W. B. Latter\footnote{Caltech, NASA Herschel
Science Center, MS 100--22,Pasadena, CA 91125} \ , S. Madden$^{12}$ \
, A. Mizuno$^{16}$, N. Mizuno$^{16}$ \ , A. Nota$^5$,
M. S. Oey\footnote{Department of Astronomy, University of Michigan,
830 Dennison Bldg., Ann Arbor, MI 48109} \ , T. Onishi$^{16}$,
R. Paladini$^{13}$, N. Panagia$^5$, P. Perez-Gonzalez$^9$,
H. Shibai$^{16}$, S. Sato$^{16}$, L. Smith\footnote{Department of
Physics and Astronomy, University College London, Gower Street, London
WC1E 6BT}~\ , L. Staveley-Smith\footnote{Australia Telescope National
Facility, CSIRO, P. O. Box 76, Epping NSW 1710, Australia} \ ,
A.G.G.M. Tielens\footnote{Kapteyn Institute, P.O. Box 800, NL-9700 AV
Groningen, Netherlands} \ , T.  Ueta\footnote{NASA Ames Research
Center/SOFIA, Mail Stop 211-3, Moffett Field, CA 94035} \ , S. Van
Dyk$^{11}$, and D. Zaritsky$^9$}

\begin{abstract}

Color--magnitude diagrams (CMDs) are presented for the {\it Spitzer}
SAGE (Surveying the Agents of a Galaxy's Evolution) survey of the
Large Magellanic Cloud (LMC). IRAC and MIPS 24 \mic \ epoch one data
are presented. These data represent the deepest, widest mid--infrared
CMDs of their kind ever produced in the LMC. Combined with the 2MASS
survey, the diagrams are used to delineate the evolved stellar
populations in the Large Magellanic Cloud as well as Galactic
foreground and extragalactic background populations.  Some 32000
evolved stars brighter than the tip of the red giant branch are
identified. Of these, approximately 17500 are classified as
oxygen--rich, 7000 carbon--rich, and another 1200 as ``extreme''
asymptotic giant branch (AGB) stars. Brighter members of the latter
group have been called ``obscured'' AGB stars in the literature owing
to their dusty circumstellar envelopes. A large number (1200) of
luminous oxygen--rich AGB stars/M supergiants are also
identified. Finally, there is strong evidence from the 24 \mic \ MIPS
channel that previously unexplored, lower luminosity oxygen--rich AGB
stars contribute significantly to the mass loss budget of the LMC (1200
such sources are identified).

\end{abstract}


\section{Introduction}

The Spitzer Space Telescope (Spitzer), with its rapid wide field
mapping capability and sensitivity, allows us to survey the thermal
emission produced by mass loss from individual evolved stars in the
nearest galaxies. \citet{mex06} presented the {\it Spitzer} SAGE
(Surveying the Agents of a Galaxy's Evolution) survey of the Large
Magellanic Cloud (LMC) along with a detailed description of its goals,
expected extent in depth and coverage and some preliminary results.
The SAGE survey is designed to enable studies of the life--cycle of
baryonic matter, as traced by dust emission, in the LMC. In this work,
we focus on one corner stone of this life--cycle, the dusty evolved
star population, which is returning matter to the interstellar
medium. Coupled with detailed star formation history (SFH) studies at
optical wavelengths, SAGE provides the opportunity to link specific
LMC populations to their dust content as a function of key parameters
such as metallicity and age. Such analyses can, in turn, be used to
validate predictive models of infrared emission in high redshift
galaxies. In the present paper, we focus on the necessary first step
of identification and characterization of the dusty evolved star
populations.

SAGE provides a global view of the dust producing stellar populations
in the LMC that will contribute to our understanding of evolved stars.
Essentially all significant, dusty, mass-losing stars in the LMC will
be detected by SAGE as it provides the deepest, widest survey in
the near-- and mid-infrared wavelengths (3--24 microns) produced to date
for this nearby galaxy. These wavelengths are particularly well
suited to the investigation of the late stages of stellar evolution
because once beyond the near--infrared (five microns), emission from
circumstellar dust can become the dominant source of emission over the
radiation from the photosphere. By eight microns, many LMC sources
show appreciable emission from dust, and the MIPS 24 micron band is
extremely sensitive to cool envelopes with excess dust emission even
for sources with little evidence of strong dust excess at shorter
wavelengths.

SAGE builds on and leverages a wealth of past observations of evolved
stars in the LMC. Optical \citep[MCPS; see][and
http://ngala.as.arizona.edu/dennis/mcsurvey.html]{zar04} and
near--infrared surveys, for example, DENIS \citep{denis} and 2MASS
\citep{skrut06}, have provided a global view of the LMC at shorter
wavelengths. See \citet{cioni06} and \citet{niko00} for LMC data from
these two surveys, respectively. The Infrared Astronomy Satellite
\citep[IRAS,][]{neug84} revolutionized the study of evolved stars by
observing objects with luminous dusty envelopes at long
wavelengths. This survey has provided samples of stars in the LMC to
be observed in greater detail up to the present, including
spectroscopic observations with {\it Spitzer}; see \citet{zijl06} and
\citet{mark05}. However, IRAS was only sensitive enough to observe the
brightest sources in the LMC \citep{schwer89} from which several
hundreds of mass--losing evolved star candidates could be identified
\citep{loup97}. The successful Midcourse Space Experiment
\citep[MSX,][]{price01} was about four times more sensitive and
provided a full LMC survey \citep{egan01}. SAGE should be roughly 1000
times more sensitive than MSX when complete \citep[][the present data
are roughly 400 times more sensitive]{mex06}. While near--infrared and
optical surveys of the LMC have identified lower mass evolved stars,
only SAGE has the depth to detect such stars at wavelengths necessary
to explicitly explore their dust properties. The ISOCAM Magellanic
Cloud Mini--Survey \citep[see][]{cioni03} has provided a preliminary
mid--infrared view of the LMC over a small field of view (0.3 square
degrees) near the depth of SAGE.

The SAGE sensitivity and areal coverage of the entire LMC will allow a
detailed quantitative derivation of the global mass loss budget from
all stellar populations when combined with existing and future
mid--infrared spectroscopic observations of the asymptotic giant
branch (AGB) stars and supergiants; for example, see \citet{vloon99},
\citet{vloon05}, \citet{mark05}, and \citet{zijl06}. Several studies
have derived detailed SFHs in the LMC from a variety of field
locations and clusters \citep{holtz99,olsen99} using principally
optical color--magnitude diagrams (CMDs) and sophisticated stellar
models and fitting techniques \citep[e.g.][]{harris01}. Recently,
\citet{cioni06} made a global SFH calculation using the 2MASS and
DENIS surveys. Their new models still have difficulties producing the
correct mix of observed carbon--rich AGB stars (C--rich or C--stars)
versus oxygen--rich AGB stars (O--rich or M stars), but their
extension to these wavelengths of the general technique is a great
advance. With SAGE, we will be able to associate the mass--loss and
chemical properties of the AGB stars in the LMC with their known
evolutionary state from existing SFH studies. This coupling can then
be used to investigate, with precision, the evolutionary status of
galaxies which lie at much larger distances through the development
and testing of detailed population synthesis models
\citep[for example, see][]{mouhlan03}.

In this paper, we continue the analysis begun by \citet{mex06} by
presenting CMDs of the survey data for the first of two epochs. We lay
the ground work for the detailed analysis to come when the full SAGE
survey is available by dissecting the observed CMDs and indicating
where various LMC and fore and background populations reside in these
diagrams.  The present paper covers approximately 49 square degrees in
the IRAC bands and MIPS 24 \mic \ band.

\section{The SAGE Catalog and Epoch One Source List}

\begin{figure}
\plotone{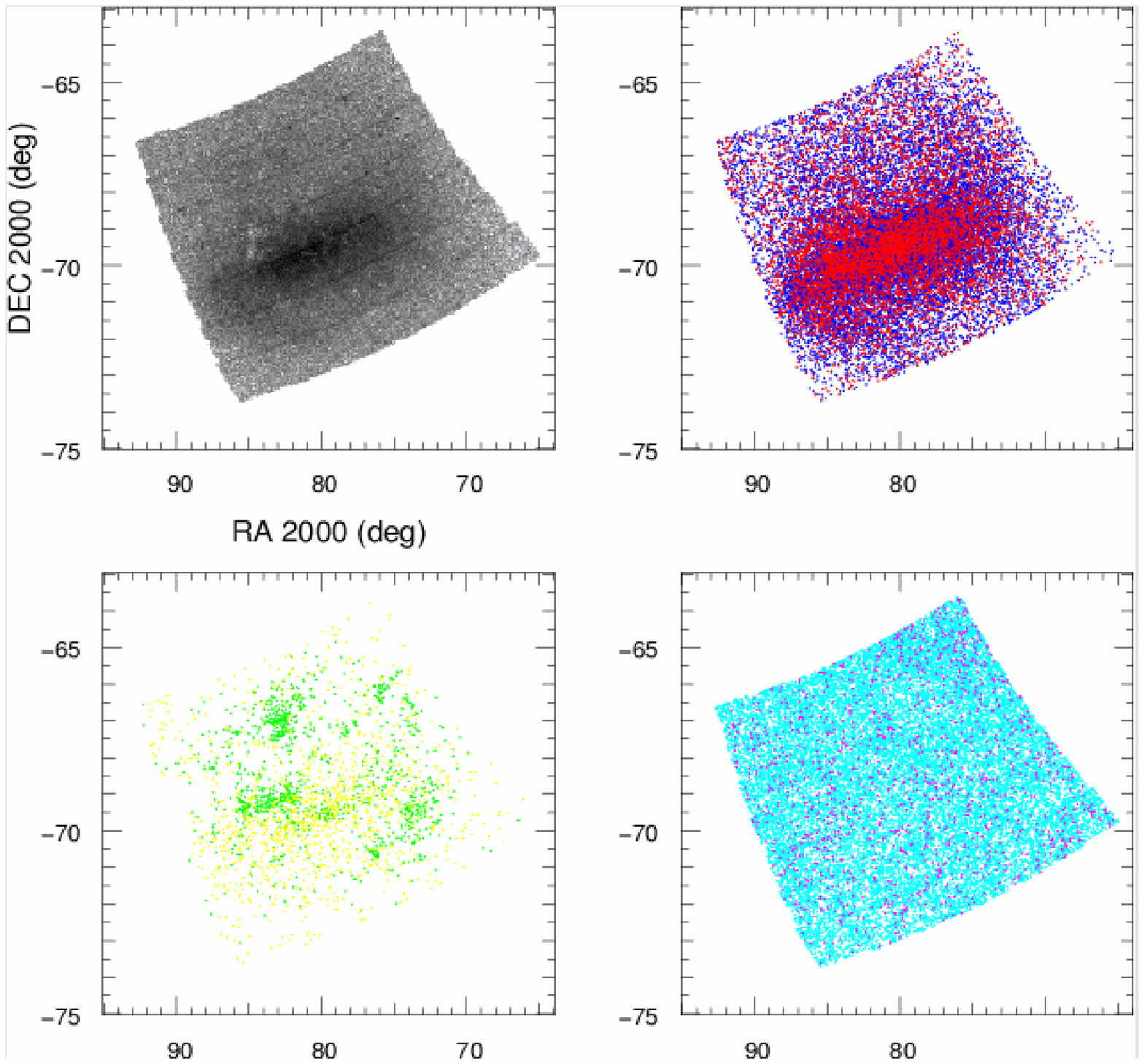} \figcaption{Positions of IRAC sources for the
epoch one dataset covering approximately 49 square degrees centered on
the Large Magellanic Cloud. The {\it upper left} panel shows the
source density in gray scale, highlighting the stellar bar. The figure
shows sources detected at $J$, [3.6], and [8.0]. The points are color
coded according to stellar or object type (see text and
Figure~\ref{j3} for details). The {\it upper right} panel shows the
distribution of carbon--rich ({\it red}) and oxygen--rich ({\it blue})
stars. The {\it lower left} panel shows the ``extreme'' asymptotic
giant branch stars ({\it yellow}) and luminous oxygen--rich stars and
M supergiants ({\it green}). The {\it lower right} panel shows two
source types which may be dominated by background galaxies: the
spatial distributions are uniform (see text). The {\it magenta} points
in this panel show objects with strong $J-$[8.0] excess while the {\it
cyan} objects have no $J-$band counterparts. 
\label{allradec}}
\end{figure}

The SAGE epoch one catalog is discussed by \citet{mex06}. The present source
list contains the first epoch data for the IRAC (3.6 \mic, 4.5 \mic,
5.8 \mic, and 8.0 \mic) bands processed by the IRAC pipeline as of May
04, 2006. This amounts to a coverage of approximately 49 square
degrees (Figure~\ref{allradec}). The current IRAC source list includes
a merged source list with the 2MASS point source catalog. Details of
the SAGE IRAC pipeline processing are given by \citet{mex06}; the
pipeline is based on the GLIMPSE pipeline \citep{ben03}. Briefly, the
IRAC data were taken such that a typical point on the sky was observed
twice within a $\sim$ one degree square ``tile.'' The frames are
analyzed with a custom DAOPHOT \citep{stet87} package developed by the
GLIMPSE team. The catalog photometry is based upon combining the
individual frame photometry (i.e., photometry is not done on the
mosaic tiles). Each tile itself overlaps at the edges with adjacent
tiles and the comparison of source photometry between tiles results in
more detections in the overlap regions. The IRAC bands and 2MASS
catalog photometry are merged in the so--called ``cross--band'' merge
step of the pipeline which is based upon positional matching
\citep[see][and references therein for details]{mex06} of the final
flux calibrated sources in each IRAC and 2MASS band. Systematic
offsets in the IRAC--2MASS merging are less than 0.3$''$.

The combined IRAC--2MASS source list analyzed in this paper contains
approximately four million sources with a 3.6 \mic \ data point (the
deepest IRAC band in terms of SAGE photometry). There are
approximately 820000 sources with $J$ and [3.6] magnitudes of which
nearly 650000 have errors less than 0.1 mag in each of $J$ and
[3.6]. An additional 10000 sources have individual $J$ and [3.6]
errors less than 0.2 mag and the remainder have errors of less than
0.3 mag. The present dataset further has 250000 8 \mic \ sources which
also have 3.6 \mic \ data (90$\%$ have [8.0] and [3.6] photometry with
errors less than 0.1 mag). The analogous numbers for $J$ and 8 \mic \
are 230000 and 91$\%$. The IRAC uncertainties are those reported by
the custom DAOPHOT package used in the pipeline and so include photon
statistics and fitting uncertainties. The 2MASS errors are taken from
the 2MASS point source catalog. The initial frames delivered from the
{\it Spitzer} Science Center (SSC) are flux calibrated (SSC pipeline
version s12.0). A network of 238 absolute calibration stars has been
developed near the area observed by SAGE (137 stars overlap the SAGE
survey area) using the identical technique by which IRAC primary
standards were established \citep{coh03}.  Comparison between
predicted and observed magnitudes for stars in this network indicate
that ensemble systematic uncertainties in any IRAC band do not exceed
the five percent level.

\begin{figure}
\plotone{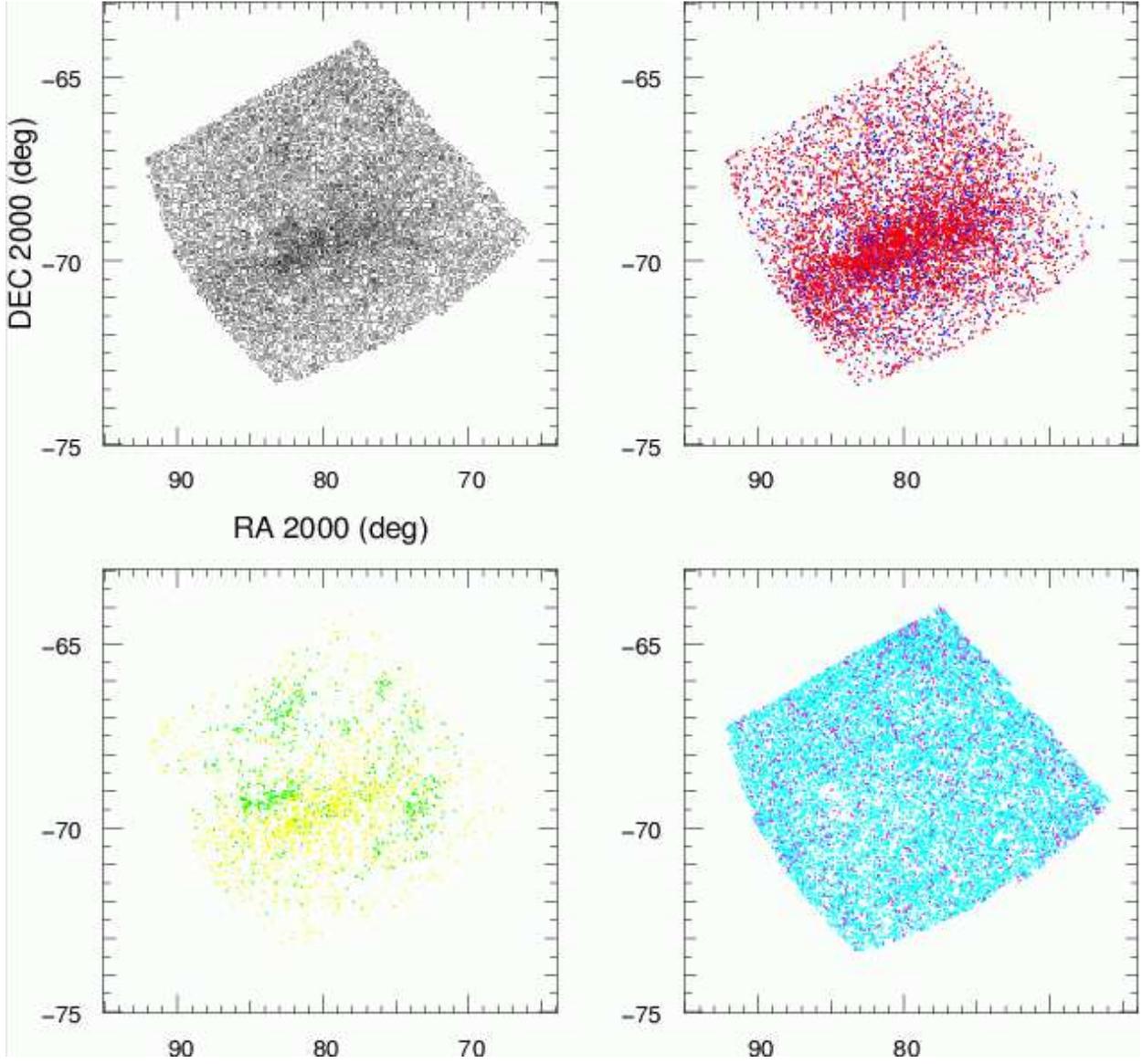} \figcaption{Same as Figure~\ref{allradec},
but for sources detected at 24 \mic \ and matched to IRAC
counterparts. \label{allradecmips}}
\end{figure}

The present source list for the MIPS 24 \mic \ band covers a similar
area as for the IRAC source list (Figure~\ref{allradecmips}). The 24
\mic \ source list (64800 sources) is for epoch one data processed as
of May, 19 2006. Details of the MIPS processing are given by
\citet{kg05} and \citet{mex06}. The MIPS data were observed in sets of
scans with each set covering a four degree by 25 arcminute strip. The
photometry was extracted via PSF fitting using the StarFinder program
\citep{diol00}.

The IRAC and MIPS data presented here were merged after assembly of
the individual catalogs by the respective pipeline teams. The IRAC
sources (those with valid 8 \mic \ magnitudes) were matched within
1$''$ of the MIPS 24 \mic \ sources, choosing the closest companion if
multiple possibilities existed.  This resulted in 27741 matches with a
an average difference in position of 0.42$''$ (and systematic offsets
in RA of $-$0.024$''$ and DEC of $-$0.004$''$) and a distribution of
offsets with one sigma standard deviation of approximately
0.25$''$. The 24 \mic \ pixel size is 2.6$''$ and the diffraction
limited $\lambda$/D is approximately 6$''$. Crowding induced errors in
the IRAC photometry and the matching of sources between IRAC and MIPS
are discussed in appendix A below.

About 42$\%$ of the MIPS sources have IRAC matches. What about the
other 58$\%$?  This number of non--matches suggests a large number of
very red sources toward the LMC. As the analysis below will suggest,
some of these may be background galaxies not detected by IRAC. Others
could be embedded sources in the LMC, and many could be slightly
extended sources which have no point--like IRAC counterpart. Detailed
analysis of the 24 \mic \ and other MIPS channels will be the subject
of a subsequent SAGE paper.

\section{Color--Magnitude Diagrams}

Figures~\ref{j3}--\ref{824} present four Hess diagrams (Hess diagrams
are a two dimensional histogram of the CMD which give the number of
stars per unit magnitude and color with the grayscale representations
graphically showing the relative source densities) and their
corresponding CMDs produced with IRAC, MIPS, and 2MASS photometry. The
Hess diagrams are binned into 200$\times$200 pixels (i.e., bins of
color and magnitude).

There is a wealth of information in these diagrams. \citet{egan01}
presented near--infrared$+$mid--infrared CMDs for the entire LMC from
the MSX satellite and 2MASS \citep{skrut06}. The MSX sensitivity is
approximately eight magnitudes brighter than SAGE, so that the present
survey will greatly extend the mid--infrared view of the LMC begun by
IRAS and MSX.

In the following, all near--infrared photometry has been taken from
the 2MASS point source catalog \citep{skrut06}. These data have been
merged with the SAGE photometry using the SAGE IRAC pipeline (see
above). The diagrams in this paper use IRAC and MIPS photometry
converted to magnitudes using the following zero points
\citep{reach05, engel06}: [3.6] zero mag $=$ 280.9 Jy,
$\lambda_{\circ}$ $=$ 3.55 \mic; [8.0] zero mag $=$ 64.13 Jy,
$\lambda_{\circ}$ $=$ 7.872 \mic, [24] zero mag $=$ 7.15 Jy,
$\lambda_{\circ}$ $=$ 23.68 \mic.

\subsection{IRAC and 2MASS Color--Magnitude Diagrams}

Earlier ``all--LMC'' surveys by IRAS and MSX had sensitivities 1000
times less than SAGE. While these surveys could detect the brightest
sources in the LMC, it has been more difficult to put them in the
overall context of the entire stellar population. In the following
sections we analize the SAGE CMDs with an emphasis on identifying the
main evolved star populations and enumerating their relative
contributions to the total source counts. This basic accounting will
help guide future detailed studies using the SAGE data base as well as
provide a comprehensive means for selecting follow--up targets for
spectroscopic study.

\subsubsection{$J-$[3.6] vs. [3.6]}

The 3.6 \mic \ channel is by far the most sensitive of the IRAC bands
and as such provides the deepest photometry to compare to 2MASS.
\cite{niko00} presented the 2MASS CMD for the LMC. Comparison of their
Figure~3 with the $J-$[3.6] $vs.$ [3.6] diagram shown in
Figure~\ref{j3} shows the same sequences with perhaps slightly more
``contrast'' in the features owing to the longer baseline in
wavelength. Starting with $J-$[3.6] $vs.$ [3.6], the first prominent
finger ($J-$[3.6] \aple 0.5) reaching to bright magnitudes
([3.6]$\sim$6) corresponds to region ``B'' of \citet{niko00}, young
A--G supergiants. There is clearly a slope to this finger indicating
predominantly an LMC population (foreground sequences appear vertical
due to the varying distance of the sources which smears out their
magnitudes but not their colors). The brightest objects in this sequence
include Galactic K and G dwarfs. To the blue and fainter of this
feature lies the OB star locus in the LMC \citep[region ``A''
of][]{niko00}.

The next sequence to the red is the vertical one reaching to bright
magnitudes. This finger consists mainly of foreground dwarfs and
giants \citep[region ``C'' of][]{niko00}. In fact, \citet{niko00} find
\apge 70$\%$ of the stars in this region of the diagram are Galactic
foreground. The importance of this contamination declines at lower
brightnesses relative to the rest of the diagram as is clearly seen in
the Hess diagram. Simple models of Galactic structure \citep[for
example,][]{blum95} predict only several late M stars in the LMC
foreground owing to their rarity.  The remaining sequences in the
$J-$[3.6] $vs.$ [3.6] diagram are predominantly LMC red giant branch
stars (RGB), AGB stars and late--type (mostly M) supergiants (SG).

\citet{cioni06} computed star formation history models of the LMC
using the 2MASS CMD and new stellar models \citep{mar03}. The
\citet{mar03} models could, for the first time, reach the observed red
colors of the C--stars and thus explain the origin of their location
in the CMD. Based on their analysis of the 2MASS CMD, \citet{cioni06}
divided the region above the tip of the RGB (TRGB) into O--rich and
C--rich zones using relations for the 2MASS $J-K$ color and $K$
magnitude. In Figures~\ref{j3}--\ref{824} the {\it blue} and {\it red}
points correspond to Cioni et al.'s {\it photometric} division of
O--rich and C--rich stars, respectively, using 2MASS colors (their
equations 5--7). The C--star and O--rich star loci are well defined
and separated in our CMDs, lending support to the \citet{cioni06}
criterion.

The {\it green} points in Figure~\ref{j3} (and following figures)
represent M SG and luminous, O--rich M stars. The locus was obtained
by using a cut in the $J-K$ $vs.$ $K$ plane which paralleled the
\citet{cioni06} relation for O--rich AGB stars, but was displaced to
the blue enough to encompass the obvious sequence in Figure~\ref{j3}
({\it left} panel) for the $J-$[3.6] $vs.$ [3.6] diagram. Cross
reference with SIMBAD (here and elsewhere in this work, SIMBAD
searches are associated with a 5$''$ search radius) shows that several
of the \citet{elias85} supergiants fall on this sequence as well as
luminous M stars identified by \citet{mass02}. \citet{niko00} confirm
the presence of known M SG on the associated $J-K$ sequence. These
luminous, late--type stars are young and thus trace regions in the LMC
of recent star formation (M SG have ages of about 10 Myr while the
most luminous AGB stars are a few hundreds of Myr). This is seen in
Figures~\ref{allradec} and \ref{allradecmips}; contrast the clumpy and
non--uniform distribution of the young stars with the more smooth
distribution of the older C--stars and O--stars. The distribution of
young M supergiants qualitatively matches the structure seen in other
tracers of star formation in the LMC; see, for example, the H$_\alpha$
map of \citet{gaus01} and the UV map of \citet{smith87}. These and
other tracers are conveniently summarized in Figure~1 of
\citet{mex06}.

The {\it yellow} points in Figure~\ref{j3} represent a group of
objects which show the effects of dusty circumstellar envelopes. These
``extreme'' AGB stars are discussed in the next section where their
$J-$[8.0] color is used to define their selection. Likewise, a
sequence which may be dominated by background galaxies is given by the
{\it magenta} points in Figure~\ref{j3}. These sources are defined by
their $J-$[8.0] color also (see below), and shown in the $J-$[3.6]
$vs.$ [3.6] CMD to demonstrate how their excess color is dominated by
longer wavelength emission.

\subsubsection{$J-$[8.0] $vs.$ [8.0]} 

C--stars have $J-$[3.6] color which extend up to about 3.0 ($J-K$
\aple 2.1); at this point, the sequence continues at nearly constant
[3.6] brightness (see Figure~\ref{j3}) to very red colors (whereas the
$K$ magnitudes break sharply to fainter brightnesses here). In the
$J-$[8.0] $vs.$ [8.0] CMD shown in Figure~\ref{j8}, the analogous
break occurs at $J-$[8.0] $\approx$ 3.5 with the [8.0] data points
trending to brighter magnitudes, but with different slope than
before. We have drawn this ``break'' in the apparent C--star relation
arbitrarily at $J-$[3.6] $=$ 3.1 and coded the redder objects with
[3.6] $\ge$ 10.5 as {\it yellow} points in
Figures~\ref{j3}--\ref{824}. We call these stars ``extreme'' AGB
stars. The brightest members of this sequence (IRAS sources) and the
red stars above it have been the subject of numerous earlier studies
to search for and characterize obscured AGB stars. See, for example,
\citet{loup97, vloon99}. While most of the objects in this sequence
are C--stars, not all of them are \citep[see, for example,
][]{zijl06}; spectroscopic confirmation is required to confidently
place any particular source in the C--star or M--star category.  A
number of investigators have also studied representatives of the
``extreme'' star sequence in detail at shorter
wavelengths. \citet{hughes90} investigated the AGB stars which span
the ``visible'' C--star locus and bluest part of the present
``extreme'' star locus ($J-$[8.0] \aple 4).

The separation of the stellar sequences becomes more apparent in the
$J-$[8.0] $vs.$ [8.0] diagram. The prominent middle vertical sequence
(``C'') now breaks into two clear sequences with foreground late K
giants further to the red than earlier G type giants.

An obvious sequence of sources becomes evident in the $J-$[8.0] $vs.$
[8.0] diagram. These points are color coded in {\it magenta}. As
pointed out by \citet{mex06}, comparison of the [8.0] number counts to
those for deep extragalactic fields \citep{faz04} indicates these are
predominantly background galaxies. The same sources are shown in the
$J-$[3.6] diagram where they merge with the bottom of the LMC
RGB. This indicates a strong 8 \mic \ excess, presumably due to
polycyclic aromatic hydrocarbons (PAHs); see, for example,
\citet{dale06}. This population is the bright component of a much
larger population which is discussed in the next section.  The number
of sources of various populations in the CMDs discussed in this
section and below are summarized in Table~1.

\subsection{The IRAC and MIPS Color--Magnitude Diagrams}

\subsubsection{[3.6]-[8.0] $vs.$ [8.0]}

The [3.6]$-$[8.0] $vs.$ [8.0] Hess diagram and CMD are shown in
Figure~\ref{38}. The LMC and foreground stars are compressed,
generally, to the blue except for the extreme AGB stars. In fact, some
of these are so extreme that they are not detected at $J$ in
2MASS. The brightest ([8.0] \aple 7), reddest ([3.6]$-$[8.0] \apge
1.5) objects (both {\it yellow} and {\it cyan}) are typically IRAS
sources and/or MSX \citep{egan01} LMC sources, a number of which have
already been observed spectroscopically with the {\it Spitzer} IRS
\citep[see also the following section]{zijl06,mark05}.

The faint cloud of objects centered at 3, 12.5 ({\it cyan} points) in
Figure~\ref{38}, is likely dominated by background galaxies. These
objects have no $J-$band counterparts and so appear to form the
fainter extension of the {\it magenta} sources discussed above (all
objects in this diagram colored {\it cyan} have no $J$ band
counterpart in the 2MASS point source catalog whether they are bright,
red AGB stars or faint red galaxy candidates). 
Young stellar objects (YSOs) have similar colors as those for
the background galaxies discussed here
\citep[see][]{whit04,mex06}. Detection of YSOs will therefore require
a careful analysis of background counts, consideration of the spatial
location (i.e. in or near young clusters) or clustering of such
objects, and most likely, follow up spectroscopy. The investigation of
YSOs in the LMC is a major focus of the SAGE team and will be
discussed in detail in future publications.

We have queried the SWIRE extragalactic survey database \citep{lon04}
to estimate the number of background sources expected.  Table~2 shows
the number of sources per square degree with 10 $<$ [8.0] $<$ 13.5 mag
and [3.6]$-$[8.0] $>$ 1.5 mag for three of the SWIRE fields and our
SAGE data. A direct comparison between SAGE and SWIRE is difficult for
several reasons. The SAGE data are not complete; thus the SAGE counts
should be considered a lower limit. The SWIRE photometry comes from
Sextractor \citep{bert96} while SAGE uses DAOPHOT. The criteria for
detecting extended objects is different for these two methods. It is
possible that slightly extended, faint, objects are extracted by
DAOPHOT as point sources. In Table~2, we compare the SAGE counts to
the SWIRE counts for which the extracted objects are classified as
point--like, indeterminate, and slightly extended in [3.6] (as
determined from the SWIRE database 3.6 \mic \ extended source flag
with values $-$1, 0, and 1 respectively).

A large fraction of the faint red objects appear to be background
galaxies. One of the SWIRE fields is plotted in
Figure~\ref{swire}. This CMD looks qualitatively the same as
Figure~\ref{38} in the region fainter than [8.0] $=$ 10 mag and for
[3.6]$-$[8.0] $>$ 1.5 mag; the distribution of magnitudes and colors
is similar including the rather sharp redward cut--off at
[3.6]$-$[8.0] $=$ 3.5 mag and the small number of sources brighter
than [8.0] $=$ 11 mag. The SAGE data have more counts per square
degree than all the SWIRE fields. This suggests some of the faint, red
objects belong to the LMC; however a definitive count of the LMC
background will require observations in fields near the LMC. Indeed,
the surface density of the putative background population towards the
LMC appears uniform ({\it magenta} and {\it cyan} points,
Figure~\ref{allradec}) compared to the LMC stellar populations.

\subsubsection{[8.0]$-$[24] $vs.$ [24]}

Moving to longer wavelengths, the [8.0]$-$[24] $vs.$ [24] diagram
(Figure~\ref{824}) further compresses the bluer stellar sequences such
that only stars with very strong emission in [24] stand out.  In
Figure~\ref{824}, {\it Spitzer} IRS photometry is over--plotted on the
SAGE data for a sample of sources investigated by \citet{mark06}. Many
of these are highly obscured objects. \citet{mark06} have classified
the objects according to the scheme of \citet{kraem02}. This scheme
identifies objects with silicate features in the circumstellar
envelope (``S'', typically O--rich stars in the LMC, but could also be
young, embedded stars) and carbonaceous features (``C'', typically
C--stars). In addition, objects can be classified as planetary nebulae
(PNe or ``P'').

The population of sources identified above with background galaxies
(same sources from the [3.6]$-$[8.0] $vs.$ [8.0] CMD with no $J-$band
counterparts, {\it cyan} points) is well differentiated from stars in
this diagram. We have argued above that most of these are probably
background galaxies (the brightest sources include known LMC objects;
see below). It is clear that our counts for [24] are not complete for
all [8.0]$-$[24] as evidenced by the sloping cut--off at the red limit
of Figure~\ref{824} for [24] $<$ 11 mag and [8.0]$-$[24] $>$ 3
mag. Nevertheless, a comparison to the SWIRE data shows that we should
expect approximately 200--400 galaxies per square degree (see Table~2)
with 7 $<$ [24] $<$ 11 mag. This compares with approximately 250 per
square degree for the SAGE data.

As mentioned in the preceding section, some LMC YSOs are expected to
be included in the SAGE source counts at all brightness levels. This
can already be verified for the brightest, reddest objects. The IRS
sources plotted in Figure~\ref{824} include four ``S'' objects which
are generally the brightest, reddest objects and lie in the part of
the CMD where massive YSOs are expected \citep{whit04,mex06}. Two of
these sources are associated with young objects and may indeed be
embedded, massive YSOs. Source MSX~LMC~1200, the brightest ``S''
source in Figure~\ref{824}, is identified as a likely compact H~II
region by \citet{vloon01}. Source MSX~LMC~1786 is identified with a
molecular cloud \citep{johan98} by \citet{egan01}. The other two ``S''
sources (MSX~LMC~906 and MSX~LMC~1436) have no further identification
apart from 2MASS catalog id's. Model colors and magnitudes for massive
YSOs \citep{whit04} also show that these objects can have the same
colors as those near the bright, red end of the ``extreme'' star
sequences in Figures~\ref{38} and \ref{824}, though such objects
should be rare even compared to AGB stars. The context, i.e. whether
an object is found in or around a region of star formation, will be
important in determining the it's precise nature (particularly those
sources which are completely obscured at shorter wavelengths).

The narrow vertical sequence at [8.0]$-$[24] $=$ $-0.2$ is composed
predominantly of stars with $J-$[8.0] $\le$ 1.0 (i.e. blue LMC stars
and foreground objects).  A sequence of points falls near [24] $=$ 10
but in between the main stellar and extragalactic loci. These sources
have blue colors at shorter wavelengths and so are not obvious on any
of the previous diagrams as they merge with the numerous LMC AGB stars
at shorter wavelengths. But the sequence stands out at 24 \mic,
trending to larger values of [8.0]$-$[24] excess (see Figure~\ref{824}
and the tip of the sequence marked by an ``F''). This sequence is
discussed further below.

\section{Discussion}

\begin{figure}
\plotone{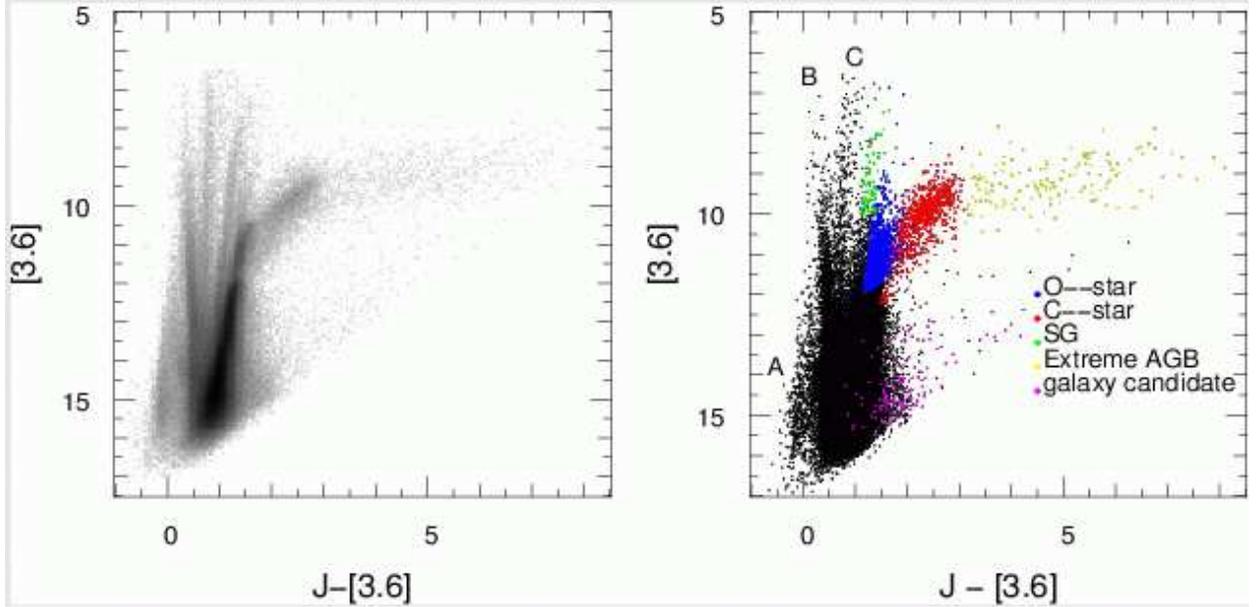} \figcaption{$J-$[3.6] $vs.$ [3.6] color--magnitude
diagram (CMD, {\it right} panel) and associated Hess diagram ({\it
left} panel). The tip of the red giant branch is located at
[3.6] $=$ 11.85 mag. The color coding is as follows. All points in the
source list are shown as {\it black} dots. {\it Blue} dots are
oxygen--rich asymptotic giant branch (AGB) stars, and {\it red} dots
are carbon--rich AGB stars. The {\it yellow} dots represent
``extreme'' AGB stars while the {\it green} dots are M supergiants and
luminous oxygen--rich stars. The {\it magenta} dots may be
predominantly background galaxies (see text). In the Hess diagram, the
red giant branch is the most densely populated region. The peak
density is approximately 2400 stars in a 0.045$\times$0.055 square
magnitude pixel and occurs at $J-$[3.6], [3.6] $=$ 0.8, 15.6. The
prominent sequences are described in the text; see also Table~1. The
features labeled as A, B, and C correspond to the same features seen
in the 2MASS CMD \citep{niko00}. For clarity, only every 10th source
is plotted in the CMD (all points are plotted in the Hess
diagram). \label{j3}}
\end{figure}

\begin{figure}
\plotone{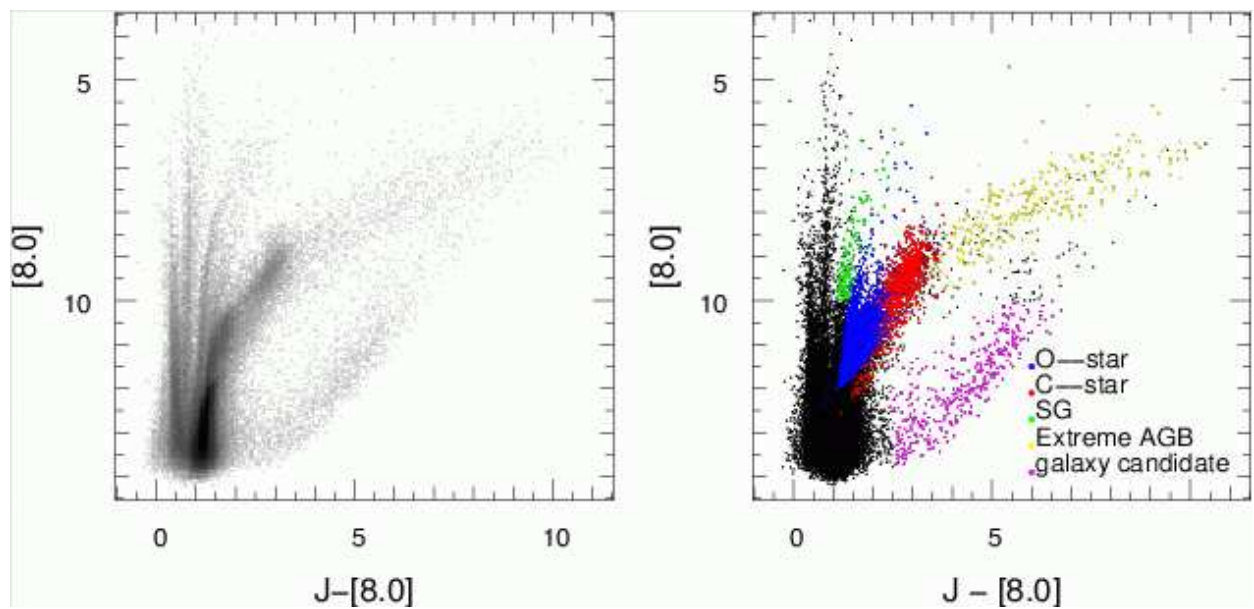} \figcaption{Same as Figure~\ref{j3}, but for the
$J-$[8.0] $vs.$ [8.0] color--magnitude diagram and corresponding Hess
diagram. For clarity, only every
5th source is plotted in the CMD.\label{j8}}
\end{figure}

\begin{figure}
\plotone{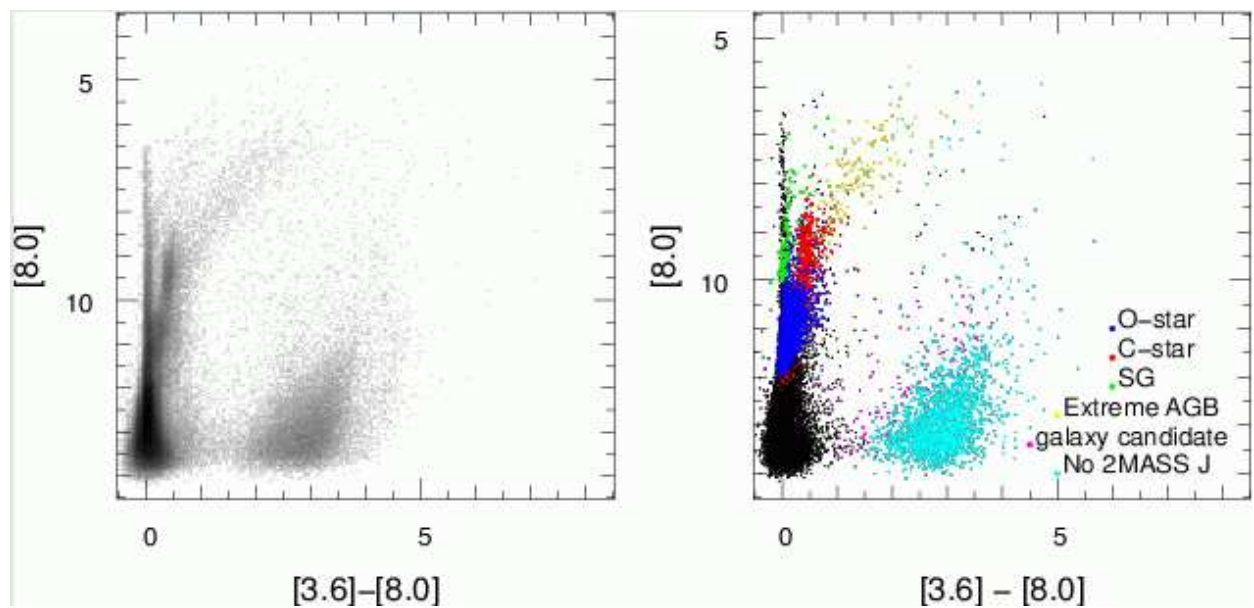} \figcaption{Same as Figure~\ref{j3}, but for the
[3.6]$-$[8.0] $vs.$ [8.0] color--magnitude diagram and Hess
diagram. Additionally, the {\it cyan} points may be dominated by
background galaxies, but which have no 2MASS $J-$band detection. The
brightest {\it cyan} points ([3.6] \aple 10 mag) are probably mostly
luminous LMC sources; see text. For clarity, only every 10th source is
plotted in the CMD. \label{38}}
\end{figure}

The Hess diagrams and CMDs presented in Figures~\ref{j3}--\ref{824}
can be used to assess the relative importance to the mass loss budget
in the LMC. Detailed model calculations will be made once the full
SAGE data set is in hand {\it and} there are appropriate template
spectra to assess the stellar envelope dust properties as a function
of location in these diagrams. The SAGE team and others have existing
and planned {\it Spitzer} and ground based spectroscopic follow--up
programs underway. Some of the brightest mass--losing sources in the
LMC have already been analyzed \citep[see, for
example,][]{zijl06}. These sources are prolific, but rare mass--losing
objects. The full impact of the less luminous stars on the AGB has yet
to be quantified.

The effect of dust in shaping the infrared CMDs is clear at a glance
from Figures~\ref{j3}--\ref{824} and considering the 2MASS $H-K$ $vs.$
$K$ CMD \citep{niko00}. In the latter, even at $Ks$ (2.1 \mic), the
tail of extreme AGB stars decreases in brightness for redder
colors. This suggests the primary effect at short wavelengths is
circumstellar extinction. In the $J-$[3.6] $vs.$ [3.6] diagram of
Figure~\ref{j3}, the extreme star branch has essentially constant
brightness. By 8 \mic, the sequence is increasing in brightness
clearly indicating that the circumstellar envelopes are exhibiting
successively more excess emission. The [8.0]$-$[24] $vs.$ [24] CMD
shows that the majority of the reddest objects are the fainter objects
we have identified with background galaxies. However, at the brightest
and reddest this sequence seems to merge with the luminous stellar
sources (i.e. the ``extreme'' AGB and SG stars). Indeed a survey of
the SIMBAD catalog shows a number of these objects to be PNe and other
bright, stellar, IRAS sources (one Wolf--Rayet star and several
objects associated with HII regions, as well as obscured AGB
stars). Of the 236 sources with [8.0]$-$[24] $>$ 3.0 mag and [24] $<$
7.0 mag, 10 are known PNe and 13 are classified as IRAS or IR sources
\citep[of which ][discuss four as obscured AGB stars]{loup97}, The
majority have no other counterparts in the SIMBAD data base. 

This general picture is confirmed by the {\it Spitzer} IRS sources
plotted in Figure~\ref{824}. The stars identified as ``C'' lie among
the ``extreme'' stars where many C--rich objects are
expected\footnote{One object, LI--LM~603, had anomalous IRAC/MIPS
colors. In this case, we substituted the IRS spectrophotometry.}. The
``S'', or silicate, sources lie among these, but also slightly to the
red and at brighter magnitudes where the M supergiants ({\it green}
points) and luminous AGB stars predominate. The IRS PNe lie at bright
magnitudes and red colors as do several massive YSOs as discussed in
\S~3.2.2. Clearly SAGE will provide a host of new mass losing objects
to follow up in detail.

While there are differences in detail, the location of the basic LMC
AGB and supergiant sequences correspond remarkably well to those
predicted from Galactic models (moved to the distance of the LMC,
distance modulus$=$18.5) for the analogous objects
\citep{wain92,coh93,coh94,coh95}. In particular, this includes the
general location, extent, and slope of the ``extreme'' AGB stars in
Figures~\ref{j3} and \ref{j8}, as well as the position of the most
luminous supergiants and O--rich objects above the extreme stars. The
predicted positions of the Galactic objects such as PNe and HII
regions are also in rough agreement with the objects identified above
in the [8.0]$-$[24] $vs. $[24] CMD (Figure~\ref{824}).

Figure~\ref{j3} shows the most dominant {\it Spitzer} population by
number is the red giants. The RGB is the most densely populated
sequence with a tip magnitude of TRGB([3.6])$=$ 11.85. There are
approximately 650000 stars on the RGB below the TRGB in the epoch one
data set. The peak number density (2400 stars in a
0.0475$\times$0.0625 square magnitude pixel) occurs at $J-$[3.6],
[3.6] $=$ 0.8, 15.6.  The RGB number density drops rapidly to a value
$\sim$ 25$\%$ of its value just to the faint side of the
tip. Approximately 74$\%$ of the stars in the $J-$[3.6] $vs.$ [3.6]
CMD above the TRGB are O--rich, C--rich, and SG types. Roughly 12$\%$
of stars above the TRGB are foreground giants and dwarfs with the
remainder being mostly blue LMC supergiants according to the
discussion in \S3.

The AGB stars and supergiants can be divided up among those with
strong mass loss (the so called ``extreme'' stars) and those whose
colors suggest weaker mass loss or less dusty envelopes. Of the
latter, most of the objects can be classified as O--rich or C--rich
based on the $J-K$ $vs.$ $K$ classification of \citet{cioni06}. The
red tail of C--stars (literally where Figure~\ref{j3} shows red
points) stretches to approximately $J-$[3.6] $=$ 3.1 where the density
of objects drops significantly toward redder colors (see
Figure~\ref{j3}). This color limit is a useful indicator of the
extreme or obscured AGB stars as it corresponds roughly to the point
in the $J-K$ $vs.$ $K$ CMD where the AGB stars become fainter at
redder color (exhibiting strong circumstellar extinction). The C/O
star ratio (not including the ``extreme'' stars) is 0.39 (see
Table~1). This value is somewhat lower than given by \citet{cioni06};
however, the data analized by \citet{cioni06} includes stars at over a
larger area than considered here and for which the number of C--stars
is significant. A full analysis of the C/O star ratio with position will
follow in a subsequent paper.

According to the models of \citet{mar03}, the position of the C--stars
in the red tail \citep[discuss this in terms of $J-K$]{mar03} is due
to cooler temperatures resulting from their molecular opacity
differences compared to the O--rich stars. The opacity difference is a
direct result of the changing molecular equilibrium as dredge up
changes the ratio of C/O (and hence the important molecular
constituents in the C--rich phase compared to the O--rich phase). The
colors are thus ``photospheric'' in nature, and not until the obvious
break in the CMDs (e.g., $J-$[3.6] $=$ 3.1) does dust significantly
affect the colors. The majority of stars in the red tail of
Figure~\ref{j3} are optically visible C--stars with objective prism
identifications given by \citet{bm90} and \citet{kdm01}. About 25$\%$
of the sources identified as C--stars here have no previous
identification in the SIMBAD database (but are 2MASS sources).

Comparison of the [3.6]$-$[8.0] $vs.$ [8.0] and [8.0]$-$[24] $vs.$
[24] CMDs suggests the relative importance of mass loss can be seen
among the different AGB stars. The {\it green} SGs lie to the blue of
the {\it yellow} extreme stars in the former diagram, but generally to
the red in the latter. The SGs thus appear to have cooler, perhaps
more extended, dust envelopes. The {\it red} ``visible'' C--stars
generally have little or no 24 \mic \ excess, whereas a subset of the
{\it blue} O--rich AGB stars exhibit increasing amounts of excess
emission and presumably mass loss. This sequence is approximately at
constant [8.0] magnitude ($\approx$ 10.5). This group of AGB stars is
well above the TRGB in the shorter wavelength CMDs and represents the
brightest tip of stars classified as O--rich just at the transition to
the ``visible'' C--star locus in the $J-$[8.0] $vs.$ [8.0] CMD
(Figure~\ref{j8}). The group of stars is clearly visible as a slight
enhancement in density in the Hess diagram (Figure~\ref{j8}) above and
to the red of the thin finger of AGB stars which rise above the TRGB
(i.e. at $J-$[8.0], [8.0] $\approx$ 1.5, 10.5).  Fainter O--rich AGB
stars are not detected at high signal--to--noise in this preliminary
data set at 24 \mic. Even so, this is striking evidence for mass loss
in lower luminosity stars than has generally been explored in the
LMC. 

Van Loon et al. (1999) derived mass loss rates from a sample of 57
stars in the LMC chosen by their infrared excess. Their analysis
suggests lower luminosity AGB stars have mass loss rates \apge
10$^{-7}$ M$_\odot$ yr$^{-1}$. The SAGE 24 \mic \ data reach somewhat
below the luminosities probed by van Loon et al., and the epoch one
data will be combined with epoch two photometry, allowing SAGE to
reach even fainter sources at 24 \mic \ (0.4 mag). The fact that the
low luminosity mass losing sources we have identified above
(Figure~\ref{824}, ``F'') span a range of 24 \mic \ excess from no
excess to $\sim$ 2 mag suggests they represent the faintest population
of dusty sources with significant mass loss (10$^{-7}$ \apge M$_\odot$
yr$^{-1}$ \apge 10$^{-8}$) in the LMC. Thus, the full SAGE catalog
should detect the all the important mass losing sources in the LMC.
Adding up the first epoch SAGE sources identified here as ``extreme''
AGB stars, SG/luminous AGB, and the present lower luminosity O--rich
AGB stars, we find approximately 10$\%$ of the evolved stars above the
TRGB are likely significant mass--loss sources.

In Figure~\ref{sed}, the average spectral energy distribution (SED) is
plotted for these low luminosity AGB sources. The sample of stars is
arbitrarily divided into three bins of varying [8.0]$-$[24] color
([8.0]$-$[24] $\le$ 0.67, 0.67 $\le$ [8.0]$-$[24] $<$ 1.34, 1.34 $\le$
[8.0]$-$[24] $<$ 2.0) to show the increasing excess emission along the
sequence at constant 8 \mic \ brightness. There are 632, 439, and 75
stars in these bins, respectively. To contribute to a given bin, a
star is required to have valid photometry for each 2MASS, IRAC, and
MIPS (24 \mic) band; hence, there are more stars plotted in
Figure~\ref{824} than represented in Figure~\ref{sed} since the former
includes objects for which there is no required $H$, [4.5], or [5.8]
photometry. 

The sources in all bins have very similar SEDs up to 8 \mic, but then
exhibit increasing 24 \mic \ flux indicative of successively more cool
dust emission which we suggest is due to increasing mass loss among
otherwise very similar objects.  Careful scrutiny of Figure~\ref{sed}
shows that the stars with the least amount of 24 \mic \ excess are the
brightest at shorter wavelengths, while those with the most excess are
fainter at the shorter wavelengths. The cool dust envelopes are likely
providing additional extinction at the shorter wavelengths. While not
shown in the figure, arbitrarily scaling the SEDs to match in the near
infrared shows that they are nearly identical until about 6.8 \mic. At
this point, a clear but small excess begins to show at 8 \mic. This
fact, and the fact that no fainter population of obvious mass losing
stars is evident in the much deeper CMD shown in Figure~\ref{j8}
provides support for our claim that SAGE will detect all the important
mass losing sources in the LMC.

Could these objects be chance alignments of ``normal'' AGB stars and
background (dusty) galaxies at 24 \mic? To estimate the possible
number of chance alignments, consider the SWIRE number counts given
above. The IRAC--24 \mic \ search radius was 1$''$. The total area
covered by {\it all} 24 \mic \ sources is then 28000 sources $\times$
the area per 24 \mic \ source $=$ 0.007 square degrees. Multiplying
this by the number of SWIRE galaxies per square degree (253) gives two
chance alignments per 28000 24 \mic \ sources).

\section{Summary}

Epoch one data for the SAGE survey of the entire LMC have been
presented for the {\it Spitzer} IRAC bands and the MIPS
24 \mic \ band. For the first time, an infrared view is presented
which places the bright IRAS and MSX sources in context. The {\it
Spitzer} (and 2MASS) CMDs are dominated by red giants (650000 red
giants to the present epoch one limit of the survey) and AGB stars.

For epoch one of the SAGE survey, we find approximately 18000
oxygen--rich AGB stars, 7000 carbon--rich stars, 1200 late--type
supergiants (or luminous oxygen--rich stars), and 1200 ``extreme'' AGB
stars (optically obscured or enshrouded AGB stars) which represent the
high mass loss rate evolved star population in LMC.  Of the
approximately 30000 evolved stars above the tip of the red giant
branch, some 10$\%$, can be readily identified with dusty
mass--losing envelopes. The C/O star ratio is approximately 40$\%$
(46$\%$ if all the ``extreme'' stars are considered as C--rich). This
is an average value which can be compared to the C/O star ratio maps
of \citet{cioni06} which give a range of approximately 0.2--0.6.

The 24 \mic \ channel of the MIPS instrument clearly shows that lower
luminosity AGB stars (which are not remarkable at IRAC wavelengths)
have excess emission indicating cool dusty envelopes and mass
loss. This population has been largely unexplored to date, and SAGE
will be uniquely powerful in allowing a quantitative assessment of its
overall impact on the LMC.

Approximately 12$\%$ of the stars above the tip of the red giant
branch are foreground dwarfs and giants. The Hess diagrams suggest the
relative contamination should be less at fainter
magnitudes. Extragalactic populations stand out in the longer
wavelength color--magnitude diagrams (those which include 8 \mic \ and
24 \mic \ data points) as objects with fainter magnitudes and red colors
(perhaps in part due to excess PAH emission at these wavelengths), but
the numbers are not large compared to the total. This background
contribution agrees well with published {\it Spitzer} galaxy counts in
other fields of view.

The authors would like to acknowledge useful discussions with Mark
Dickinson about SWIRE and the background galaxy counts. The plots and
analysis in this article made use of the Yorick programing
language. This article made use of the SIMBAD database at the CDS, and
this research has been funded by NASA/{\it SST} grant 1275598 and NASA
grant NAG5--12595. The authors would also like to acknowledge the
comments of an anonymous referee regarding stellar crowding the
relation between IRAC and MIPS sources. Addressing these comments has
strengthened our results.

\section{Appendix}

Given the difference in IRAC and MIPS point spread function (PSF) size
(2$''$ and 6$''$, respectively), the issue of source matching between
the two and the impact of crowding on the color--magnitude diagrams
(CMD) needs to be addressed. In \S2, we described the matching of MIPS
and IRAC sources within a 1$''$ radius. In the case of multiple
sources, the closer source is assumed to be the appropriate
match. Depending on the level of crowding of IRAC sources within the
MIPS PSF, this matching technique could associate a fainter IRAC
source with the sum of multiple MIPS 24 \mic \ sources leading to
spurious red colors (for example in the finger of low luminosity
oxygen rich asymptotic giant branch, AGB, stars discussed in \S4).

In order to quantify the effect of the multiple IRAC sources within
the MIPS PSF we re--ran the matching of IRAC and MIPS sources with a
6$''$ radius (i.e, twice the size of the MIPS PSF full width at half
maximum). This matching resulted in a total of approximately 30000
sources (10$\%$ more than the previous matching discussed in \S2). The
distribution of offsets was similar to the original with a tail to
larger matches out to $\sim$ 6$''$. The mean offset changed from 0.4$''$
to 0.7$''$ and the systematic offsets in RA and Dec between the two
datasets remained much less than 0.1$''$ as before.

Two MIPS sources had as many as five matches within this new large
radius, four MIPS sources had four matches, 41 MIPS sources were
matched with three IRAC sources, and 1058 MIPS sources had two IRAC
matches. The ratio of the sum of 8 \mic \ flux due to all sources in
the 6$''$ radius to that of the chosen match (closest source) was less
than 1.5 in more than half of all of the multiple source cases while
248 matches had flux ratios greater than 2.0. The number of double
matches is sufficient enough to be interesting since some of the
sequences identified in the CMDs above had a similar number of stars
identified.

To assess the impact of the multiple sources, they are overplotted on
the [8.0]$-$[24] $vs$ [24] color--magnitude diagram (CMD) in
Figure~\ref{match}. It is immediately clear that the multiply matched
sources sample the CMD uniformly. Because their number is only 3.7
$\%$ percent of the total, we conclude that they have no significant
impact on the sequences of stars in the CMD. The bulk of multiple
sources lie in the region of highest source density in the ``bar''
feature of the LMC. Plotting the CMD for sources in the less crowded
region above the bar shows that all the prominent features of the CMD
described in the text remain.

\begin{figure}
\plotone{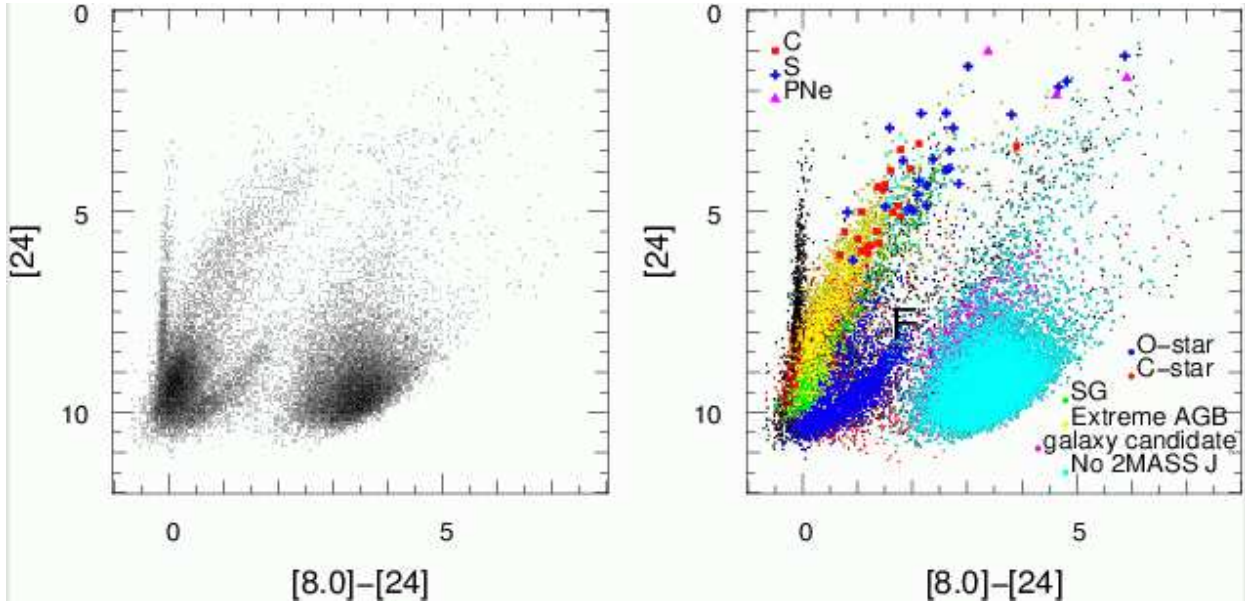} \figcaption{Same as Figure~\ref{38}, but for the
[8.0]$-$[24] $vs.$ [24] color--magnitude diagram and Hess diagram. The
brightest {\it cyan} points ([24] \aple 7) correspond to luminous LMC
stars and other objects such as planetary nebulae and young
emission--line sources.  A number of the {\it cyan} objects lie at the
tip of the {\it yellow} extreme asymptotic giant branch (AGB) star
sequence and thus are likely luminous AGB stars that are completely
obscured at shorter wavelengths. This picture is confirmed by the
plotted {\it Spitzer} IRS spectroscopy sources, ``C'', ``S'', and
``PNe'' which indicate carbonaceous, silicate, and PNe features,
respectively (see text for details).  As in Figure~\ref{38}, most of
the {\it cyan} points ([24] \apge 7) are probably background galaxies
(see text and Figure~\ref{allradecmips}). The sequence of oxygen--rich
(O--rich) AGB stars ({\it blue} dots) include a prominent ``finger''
reaching to [24], [8.0]$-$[24] $\sim$ 8.5, 2.0. These are O--rich
stars at nearly constant [8.0] magnitude (10.5, see Figure~\ref{j8}
and \ref{sed}) at the tip of the AGB which appear to have increasing
amounts of mass--loss (see text). The tip of this finger is indicated
with an ``F'' in the plot.\label{824}}
\end{figure}

\begin{figure}
\plotone{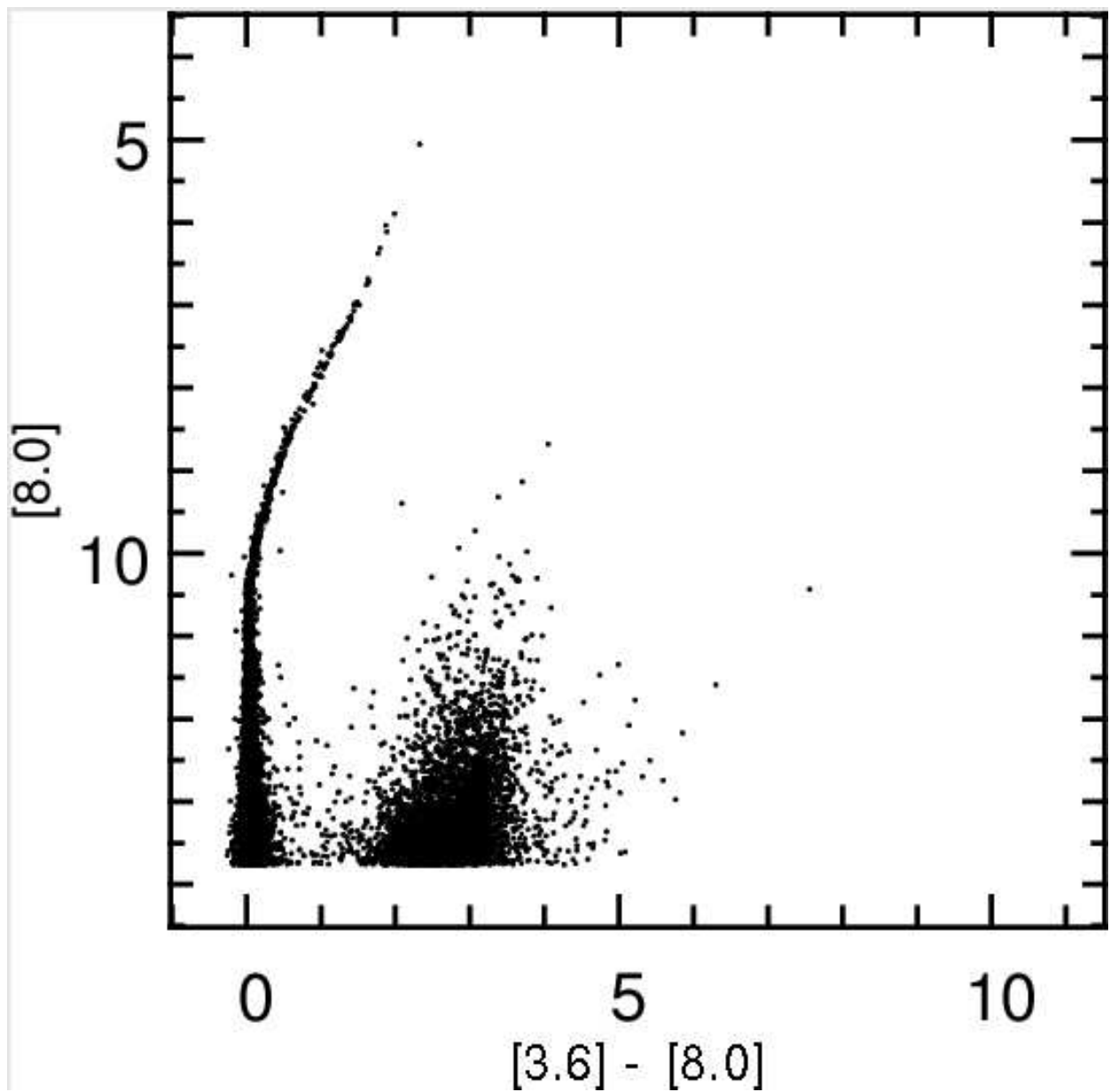} \figcaption{The [3.6]$-$[8.0] $vs.$ [8.0]
color--magnitude diagram for the SWIRE survey toward the ELAIS N1
field which covers nine square degrees. The distribution and number
(see text and Table~2) of faint, red galaxies are similar to that for
the SAGE LMC survey (Figure~\ref{38}). The curved blue sequence is
presumably due to non--linearity in the [3.6] counts due to the 30s
SWIRE exposures (SAGE uses 0.6s exposures to obtain linear data in
this range). The SWIRE sources are limited to the SAGE brightness 
limit in Figure~\ref{38}.\label{swire}}
\end{figure}

\begin{figure}
\plotone{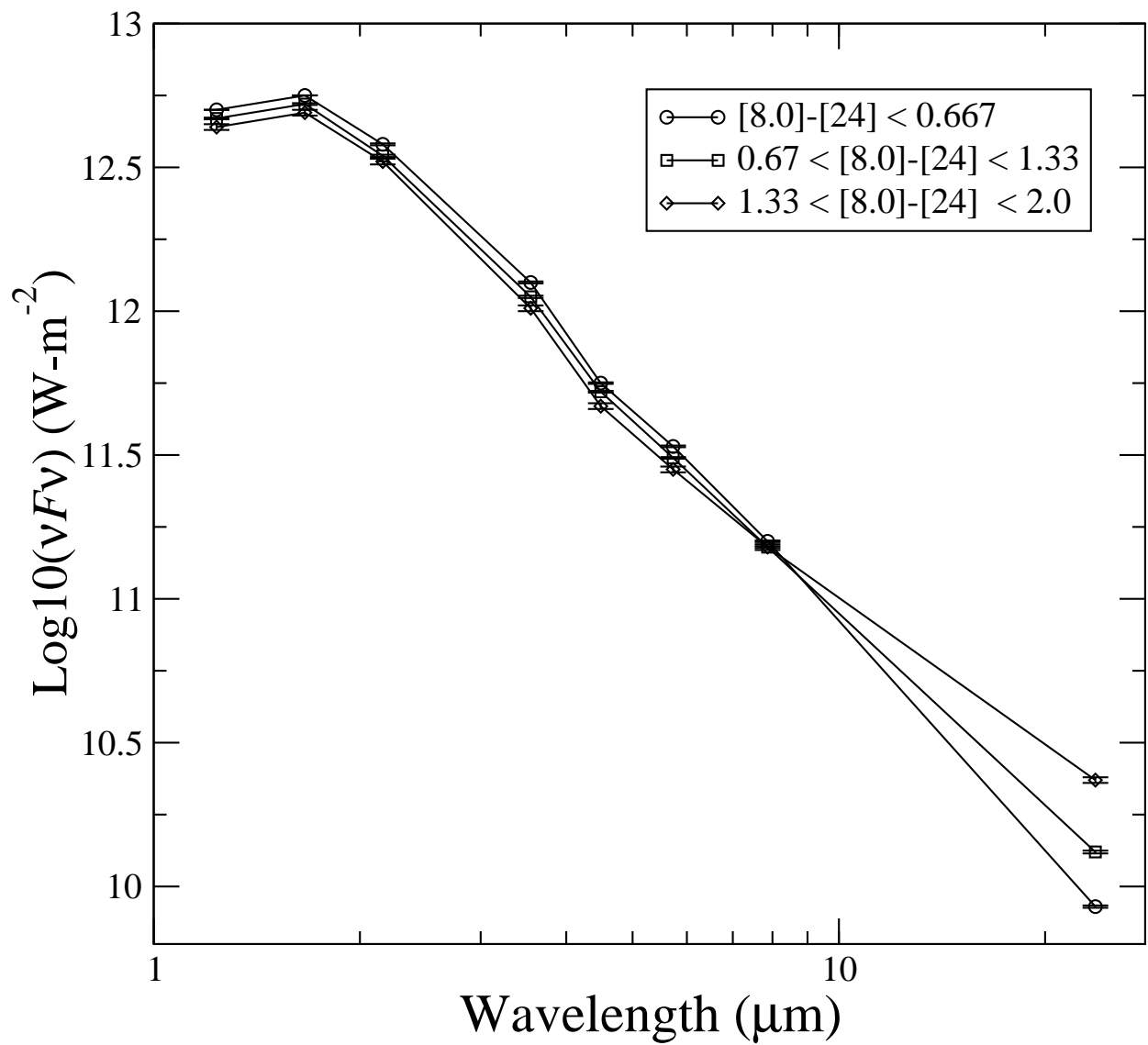} \figcaption{Spectral energy distributions (SED)
for the average source in each of three [8.0]$-$[24] color bins as
indicated in the plot. The SEDs are very similar until 8 \mic. The 24
\mic \ flux increases suggesting a sequence of stars at the tip of the
AGB (see text and Figures~\ref{j3} and \ref{j8}) exhibiting increasing
amounts of mass loss from a ``bare'' AGB star ({\it open circles}) to
an AGB star with a cool circumstellar envelope ({\it open
diamonds}). The error bars represent the error in the mean for the
average photometry in each color bin constructed from 632, 439, and
75 stars (see text). \label{sed}}
\end{figure}

\begin{figure}
\plotone{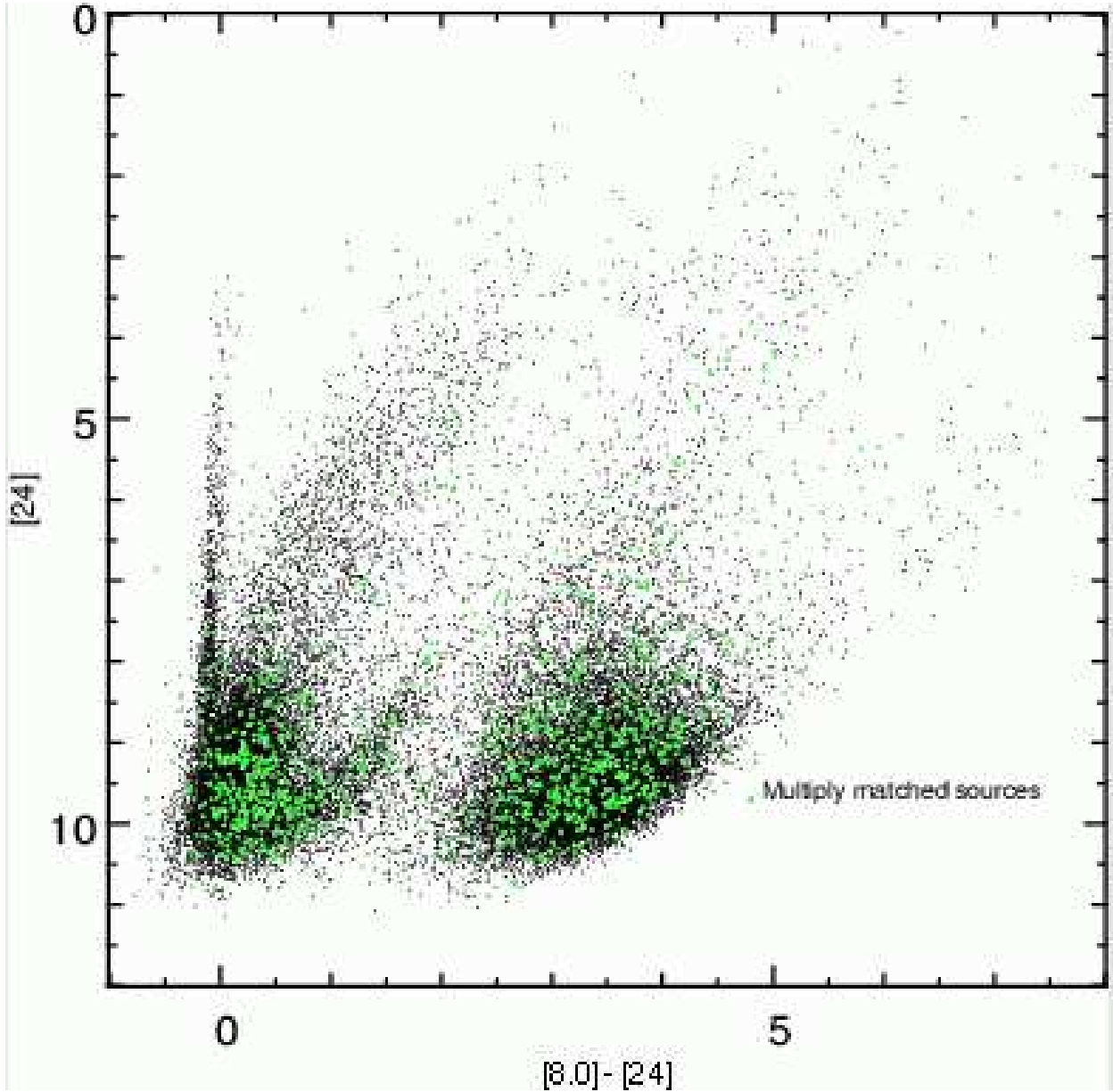} \figcaption{The [8.0]$-$[24] $vs$ [24]
color--magnitude diagram (CMD). This is the same as Figure~\ref{824},
except for an IRAC--MIPS matching radius of 6$''$. A few percent of
all the matches have multiple IRAC sources within the larger radius
used to find a MIPS 24 \mic \ source ({\it green} dots). The majority
of multiples are double with about 50 triple matches, four quadruple,
and two quintuple. The distribution of matches suggests the stellar
sources and background galaxies are similarly affected in the CMD, and
the total number of multiply matched sources for any sequence in the
CMD is small: there are 30000 {\it black} points in this diagram and
1103 {\it green} points. See the {\it appendix} for
details.\label{match}}
\end{figure}

\begin{figure}
\plotone{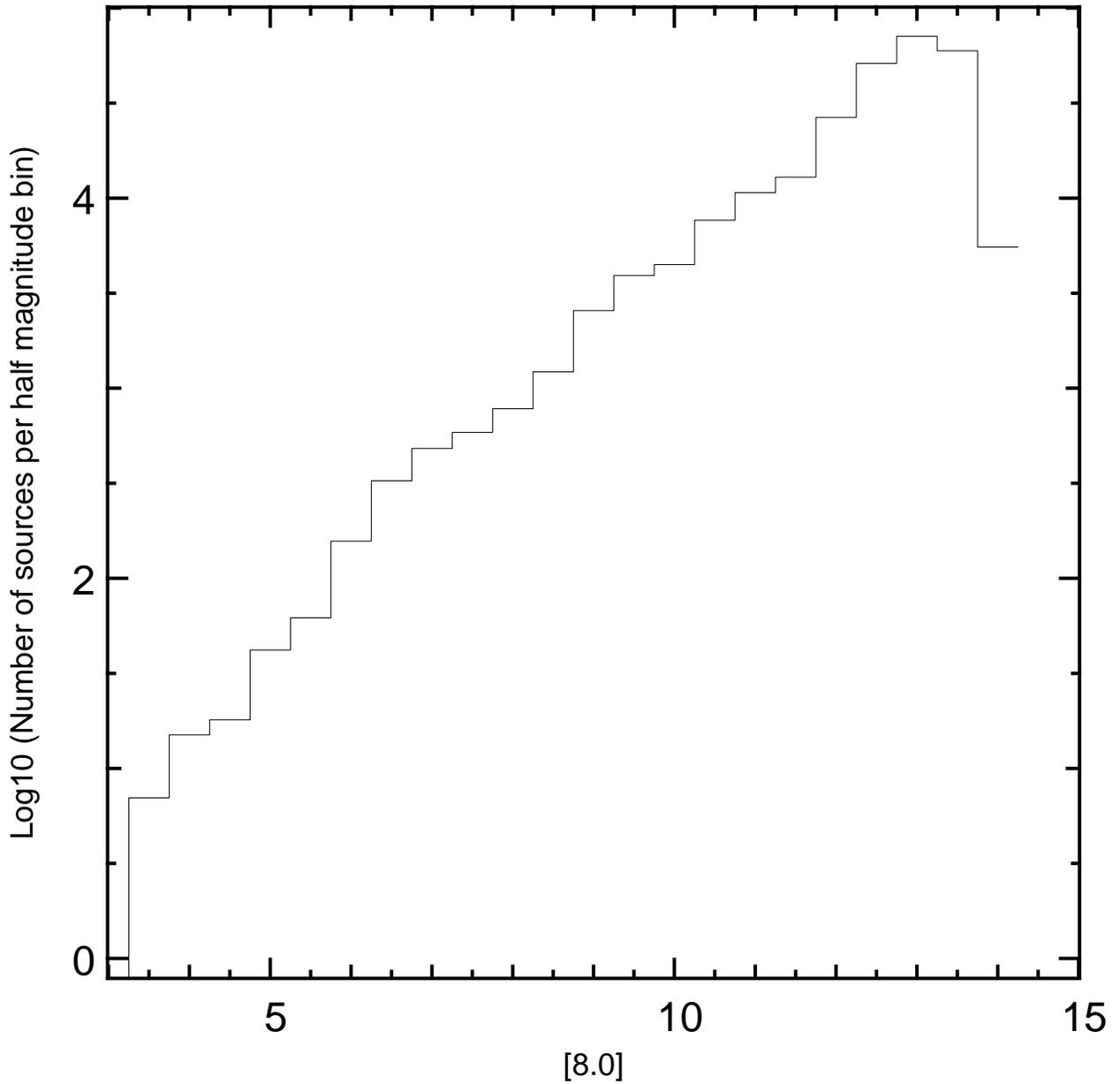} \figcaption{The 8 \mic \ luminosity function for all
8 \mic \ sources in the present epoch 1 dataset. This luminosity
function is used to show that the data in Figures~\ref{824} and
\ref{match} are not significantly affected by point source
crowding. See the {\it appendix} for details.\label{lf8}}
\end{figure}

This conclusion is based on the assumption that the IRAC 8 \mic \
counts are complete for the magnitude range considered. This is indeed
the case. Figure~\ref{lf8} shows the 8 \mic \ luminosity function for
{\it all} the IRAC 8 \mic \ sources in the SAGE epoch 1 data presented
here (approximately 260000) sources. The maximum source density is
approximately 45000 per square degree and the counts are rising to
[8.0] $=$ 13 mag (about two full magnitudes below the faintest stellar
sources in Figure~\ref{824}). The 8 \mic \ crowding limit can be
explicitly calculated following the formalism of \citet{olsen03}. The
8 \mic \ maximum surface density results in a 17 magnitude per square
arcsecond surface brightness for the luminosity function shown in
Figure~\ref{lf8}. Extrapolating the source counts to 38th magnitude
results in a surface brightness of 16.6 magnitudes per square
arcsecond. Following the prescription of \citet{olsen03}, this latter
surface brightness results in crowding induced errors of only 2$\%$ in
the [8.0] photometry to [8.0] $=$ 10.5 mag for a 2$''$ IRAC PSF. Thus
the 8 \mic \ counts should be nearly 100 $\%$ complete at the
magnitude level represented in Figure~\ref{824}, and the chance of
associating a spurious faint 8 \mic \ source with multiple 24 \mic \
sources in the MIPS beam is negligible.

\newpage


\newpage 


\begin{deluxetable}{lcccc}
\tablecaption{{\it Spitzer} SAGE Stellar Populations}
\tabletypesize{\scriptsize}
\tablehead{
\colhead{Population} &
\colhead{Diagram} &
\colhead{Number of Sources} &
\colhead{$\%$ of Total\tablenotemark{a}} &
\colhead{Figure Color\tablenotemark{b}} 
}
\startdata
Point sources\tablenotemark{c} with both $J$ and [3.6] & $J-$[3.6] $vs.$ [3.6]
 & 821305 & 100 & black\\
Red giant branch stars & $J-$[3.6] $vs.$ [3.6] & 650000 & 79 \\
Point sources brighter than TRGB\tablenotemark{d} & $J-$[3.6] $vs.$ [3.6] & 
42897 & 5 & \\
Carbon--rich AGB stars\tablenotemark{e} & $J-$[3.6] $vs.$ [3.6] & 6935 & 1 & 
red\\
Oxygen--rich AGB stars\tablenotemark{e} & $J-$[3.6] $vs.$ [3.6] & 17875&2 &blue
 \\
Supergiant/luminous Oxygen rich AGB\tablenotemark{f} & $J-$[3.6] $vs.$ [3.6]
 & 1240 & 0.1 & green\\
Extreme AGB stars\tablenotemark{g} & $J-$[8.0] $vs.$ [8.0] & 1228&0.1 & yellow 
\\
Galactic foreground brighter than TRGB & $J-$[3.6] $vs.$ [3.6] & 4995 & 1 & \\
Background galaxy candidates I & $J-$[8.0] $vs.$ [8.0] & 1887 &1 & magenta\\
Background galaxy candidates II\tablenotemark{h} & [3.6]$-$[8.0] $vs.$ [8.0] &
 17735&7 &cyan \\
24 \mic \ sources with IRAC counterparts & [8.0]$-$[24] $vs$ [24] & 27741 &100
 &black\\
Faint Oxygen--rich AGB stars with mass loss\tablenotemark{i} & [8.0]$-$[24]
 $vs.$ [24] & 1158 &4 & blue
\enddata

\tablenotetext{a}{Percentage of objects classified compared to the
total number for objects detected in the corresponding diagram (e.g.,
in $J$ and [3.6] or [3.6] and [8.0].}

\tablenotetext{b}{The color corresponding to the object type in
Figures~\ref{j3}--\ref{824}.}

\tablenotetext{c}{For the 49 square degree area shown in
Figure~\ref{allradec}}

\tablenotetext{d}{Tip of the Red Giant Branch, [3.6] $=$ 11.85; see
Figure~\ref{j3}.}

\tablenotetext{e}{Based on a photometric criterion presented by
\citet{cioni06}; see text.}

\tablenotetext{f}{See text for definition; includes arbitrary
criterion that [3.6] $\le$ 10.0 mag.}

\tablenotetext{g}{These are the so--called ``obscured'' AGB
stars. Defined here as AGB stars with $J-$[3.6] $>$ 3.1. The AGB
changes character here with negative, flat, and positive slope for
$J-K$ $vs.$ $K$, $J-$[3.6] $vs.$ [3.6], and $J-$[8.0] $vs.$ [8.0],
respectively, indicating the relative importance of dust extinction
and emission versus wavelength.}

\tablenotetext{h}{These objects have no $J-$band counterparts. The
brightest sources in this group are luminous objects in the LMC; see
text.}

\tablenotetext{i}{This sub--set of mass--losing sources is at the tip
of the oxygen rich AGB (i.e. above the TRGB) in Figure~\ref{j8} and
have [8.0]$=$10.5 $\pm$ 0.3. This sequence is indicated in
Figure~\ref{824} by the letter ``F''. The [24] detections are not
complete at this brightness level.}

\end{deluxetable}

\newpage

\begin{deluxetable}{lrc}
\tablecaption{{\it Background Galaxy Counts}}
\tablehead{
\colhead{Field\tablenotemark{a}} &
\colhead{[8.0]\tablenotemark{b} Counts deg$^{-2}$} &
\colhead{[24]\tablenotemark{c} Counts deg$^{-2}$} 
}
\startdata
SWIRE ELAIS N1 & 288 & 372 \\
SWIRE ELAIS N2 & 153 & 160 \\
SWIRE ELAIS S1 & 183 & 224 \\
SWIRE XMM & 231 & 299 \\
SWIRE Chandra DFS & 214 & 278 \\
SWIRE Lockman Hole & 247 & 321 \\
SAGE & 400 & 253
\enddata

\tablenotetext{a}{The SWIRE fields cover 9, 5, 7, 9, 8, and 11 square
degrees for N1, N2 (MIPS field is 4 square degrees), S1, XMM, Chandra,
and Lockman, respectively. The SAGE IRAC and MIPS data cover
approximately 49 square degrees.}

\tablenotetext{b}{Source counts for 10 $<$ [8.0] $<$ 13.5 and
[3.6]$-$[8.0] $>$ 1.5. The SWIRE counts are for point--like,
indeterminate, and {\it possibly} extended sources. The SAGE data are
all extracted as point sources, but faint, slightly extended objects
can not be ruled out.}

\tablenotetext{c}{Source counts for 7 $<$ [24] $<$ 11 and
[8.0]$-$[24] $>$ 1.5. See text note $b$.}

\end{deluxetable}


\newpage

\begin{thebibliography}{dummy}

\bibitem[Benjamin et al.(2003)]{ben03} Benjamin, R.A., et al. 2003,
\pasp, 115, 953

\bibitem[Bertin \& Arnouts(1996)]{bert96} Bertin, E., \& Arnouts, S.\
1996, \aaps, 117, 393

\bibitem[Blanco \& McCarthy(1990)]{bm90} Blanco, V.~M., \& McCarthy,
M.~F.\ 1990, \aj, 100, 674

\bibitem[Blum et al.(1995)]{blum95} Blum, R.~D., Carr, J.~S.,
Sellgren, K., \& Terndrup, D.~M.\ 1995, \apj, 449, 623

\bibitem[Cioni et al.(2003)]{cioni03} Cioni, M.-R.~L., et
al.\ 2003, \aap, 406, 51

\bibitem[Cioni et al.(2006)]{cioni06} Cioni, M.-R.~L., Girardi, 
L., Marigo, P., \& Habing, H.~J.\ 2006, \aap, 448, 77

\bibitem[Cohen(1993)]{coh93} Cohen, M. 1993, \aj, 105, 1860

\bibitem[Cohen(1994)]{coh94} Cohen, M. 1994, \aj, 107, 582

\bibitem[Cohen(1995)]{coh95} Cohen, M. 1995, \aj, 444, 874

\bibitem[Cohen et al.(2003)]{coh03} Cohen, M., Megeath, T.G.,
Hammersley, P.L., Martin---Luis, F., \& Stauffer, J. 2003, \aj, 125,
2645

\bibitem[Dale et al.(2005)]{dale06} Dale, D.~A., et al.\ 2005, \apj,
633, 857

\bibitem[Diolaiti et al.(2000)]{diol00} Diolaiti, E., Bendinelli, O.,
Bonaccini, D., Close, L., Currie, D., \& Parmeggiani, G.\ 2000, \aaps,
147, 335

\bibitem[Egan et al.(2001)]{egan01} Egan, M.~P., Van Dyk, S.~D., \&
Price, S.~D.\ 2001, \aj, 122, 1844

\bibitem[Elias, Frogel, \& Humphreys(1985)Elias et
al.(1985)]{elias85} Elias, J. H., Frogel, J. A., \& Humphreys,
R. M. 1985, \apjs, 57, 91

\bibitem[Engelbracht et al.(2006)]{engel06} Engelbracht, C. et
al. 2006, in preparation

\bibitem[Epchtein et al.(1994)]{denis} Epchtein, N., et al.\ 1994,
\apss, 217, 3

\bibitem[Fazio et al.(2004)]{faz04} Fazio, G.~G., et al.\ 
2004, \apjs, 154, 39

\bibitem[Gaustad et al.(2001)]{gaus01} Gaustad, J.~E., McCullough,
P.~R., Rosing, W., \& Van Buren, D.\ 2001, \pasp, 113, 1326

\bibitem[Gordon et al.(2005)]{kg05} Gordon, K., et al. 2005, \pasp,
117, 503

\bibitem[Harris \& Zaritsky(2001)]{harris01} Harris, J., \& 
Zaritsky, D.\ 2001, \apjs, 136, 25 

\bibitem[Holtzman et al.(1999)]{holtz99} Holtzman, J. et al. 1999,
\aj, 118, 2262

\bibitem[Hughes \& Wood(1990)]{hughes90} Hughes, S.~M.~G., \& Wood,
P.~R.\ 1990, \aj, 99, 784

\bibitem[Johansson et al.(1998)]{johan98} Johansson, L.~E.~B., et al.\
1998, \aap, 331, 857

\bibitem[Kontizas et al.(2001)]{kdm01} Kontizas, E., Dapergolas, A.,
Morgan, D.~H., \& Kontizas, M.\ 2001, \aap, 369, 932

\bibitem[Kraemer et al.(2002)]{kraem02} Kraemer, K.~E., Sloan, 
G.~C., Price, S.~D., \& Walker, H.~J.\ 2002, \apjs, 140, 389 

\bibitem[Lonsdale et al.(2004)]{lon04} Lonsdale, C., et al.\ 
2004, \apjs, 154, 54

\bibitem[Loup et al.(1997)]{loup97} Loup, C., Zijlstra, A.~A., Waters,
L.~B.~F.~M., \& Groenewegen, M.~A.~T.\ 1997, \aaps, 125, 419

\bibitem[Marigo et al.(2003)]{mar03} Marigo, P., Girardi, L., \&
Chiosi, C.\ 2003, \aap, 403, 225

\bibitem[Markwick--Kemper et al.(2005)]{mark05} Markwick--Kemper, F.,
Leisenring, J., Meixner, M., van Dyk, S., \& Szczerba, R.\ 2005, IAU
Symposium, 231, 134

\bibitem[Markwick--Kemper et al.(2006)]{mark06} Markwick--Kemper, F.,
Leisenring, J., Meixner M., Van Dyk S. D., Speck, A.K., Dijkstra, C.,
Szczerba R., Cami, J., \& Ueta T. 2006, in preparation.

\bibitem[Massey(2002)]{mass02} Massey, P.\ 2002, \apjs, 141, 81

\bibitem[Meixner et al.(2006)]{mex06} Meixner, M. et al. 2006,
submitted to the \aj

\bibitem[Mouhcine \& Lan{\c c}on(2003)]{mouhlan03} Mouhcine, M., 
\& Lan{\c c}on, A.\ 2003, \mnras, 338, 572 

\bibitem[Neugebauer et al.(1984)]{neug84} Neugebauer, G., et al.\
1984, \apjl, 278, L1

\bibitem[Nikolaev \& Weinberg(2000)]{niko00} Nikolaev, S., \& 
Weinberg, M.~D.\ 2000, \apj, 542, 804

\bibitem[Olsen(1999)]{olsen99} Olsen, K.~A.~G.\ 1999, \aj, 117, 2244

\bibitem[Olsen et al.(2003)]{olsen03} Olsen, K.~A.~G., Blum, R.~D., \&
Rigaut, F.\ 2003, \aj, 126, 452

\bibitem[Price et al.(2001)]{price01} Price, S.~D., Egan, 
M.~P., Carey, S.~J., Mizuno, D.~R., \& Kuchar, T.~A.\ 2001, \aj, 121, 2819 

\bibitem[Reach et al.(2005)]{reach05} Reach, W.~T., et al.\ 
2005, \pasp, 117, 978

\bibitem[Schwering(1989)]{schwer89} Schwering, P.~B.~W.\
1989, \aaps, 79, 105

\bibitem[Smith et al.(1987)]{smith87} Smith, A.~M., Cornett, 
R.~H., \& Hill, R.~S.\ 1987, \apj, 320, 609 

\bibitem[Stetson(1987)]{stet87} Stetson, P.~B.\ 1987, \pasp, 99, 191

\bibitem[Skrutskie et al.(2006)]{skrut06} Skrutskie, M.~F., et 
al.\ 2006, \aj, 131, 1163 

\bibitem[van Loon et al.(1999)]{vloon99} van Loon, J.~T., Groenewegen,
M.~A.~T., de Koter, A., Trams, N.~R., Waters, L.~B.~F.~M., Zijlstra,
A.~A., Whitelock, P.~A., \& Loup, C.\ 1999, \aap, 351, 559

\bibitem[van Loon et al.(2001)]{vloon01} van Loon, J.~T., Zijlstra,
A.~A., Bujarrabal, V., \& Nyman, L.-{\AA}.\ 2001, \aap, 368, 950

\bibitem[van Loon et al.(2005)]{vloon05} van Loon, J.~T., 
Cioni, M.-R.~L., Zijlstra, A.~A., \& Loup, C.\ 2005, \aap, 438, 273

\bibitem[Wainscoat et al.(1992)]{wain92} Wainscoat, R.~J., Cohen, M.,
Volk, K., Walker, H.~J., \& Schwartz, D.~E.\ 1992, \apjs, 83, 111

\bibitem[Whitney et al.(2004)]{whit04} Whitney, B.~A., Indebetouw, R.,
Bjorkman, J.~E., \& Wood, K.\ 2004, \apj, 617, 1177

\bibitem[Zaritsky et al.(2004)]{zar04} Zaritsky, D., Harris, J.,
Thompson, I.B. \& Grebel, E.K. 2004, \aj, 128, 1606

\bibitem[Zijlstra et al.(2006)]{zijl06} Zijlstra, A. A., et al. 2006,
astroph/0602531

\end{thebibliography}
\end{document}